\documentstyle[aps,prc]{revtex}

\newcommand{\be}{\begin{equation}}
\newcommand{\ee}{\end{equation}}
\newcommand{\ba}{\begin{array}}
\newcommand{\ea}{\end{array}}
\newcommand{\bn}{\begin{eqnarray}}
\newcommand{\en}{\end{eqnarray}}
\newcommand{\bnll}[1]{\begin{mathletters}\label{#1}\begin{eqnarray}}
\newcommand{\enll}{\end{eqnarray}\end{mathletters}}
\newcommand{\bnl}{\begin{mathletters}\begin{eqnarray}}
\newcommand{\enl}{\end{eqnarray}\end{mathletters}}
\newcommand{\bml}{\begin{mathletters}}
\newcommand{\eml}{\end{mathletters}}
\newcommand{\bc}{\begin{center}}
\newcommand{\ec}{\end{center}}
\newcommand{\bi}{\begin{itemize}}
\newcommand{\ei}{\end{itemize}}

\newcommand{\demi}{\frac{1}{2}}
\newcommand{\thalf}{{\textstyle{\demi}}}

\tighten
\begin{document}

\draft
\twocolumn[\columnwidth\textwidth\csname@twocolumnfalse\endcsname
\title{Generalization of the Bloch-Messiah-Zumino theorem}

\author{J. Dobaczewski}

\address{
Institute of Theoretical Physics, Warsaw University,
    Ho\.za 69,  PL-00681, Warsaw, Poland}

\maketitle

\begin{abstract}
It is shown how to construct a basis in which two arbitrary complex
antisymmetric matrices $C$ and $C'$ acquire simultaneously
canonical forms. The present construction is not restricted by any
conditions on properties of the $C^+C'$ matrix. Canonical bases
pertaining to the generator-coordinate-method treatment of
many-fermion systems are discussed.
\end{abstract}


\addvspace{5mm}]

\narrowtext

The Bardeen-Cooper-Schrieffer
(BCS) pairing theory \cite{[Bar57]}, and its generalization by
Bogolyubov \cite{[Bog]},
are based on fermion wave functions that have the form of
fermion-pair condensates, i.e.,
\be\label{eq201}
|C\rangle =
\exp\left(\thalf\sum_{mn}C^{\displaystyle\ast}_{mn}a^+_m a^+_n\right)|0\rangle,
\ee
where $a^+_m$ are the fermion creation operators, $|0\rangle$ is the
fermion vacuum, and $C_{mn}$ is an antisymmetric complex
matrix. Up to a unitary transformation of the single-particle
basis,
\be\label{eq202}
\bar{a}^+_m = \sum_{m'}U^{\displaystyle\ast}_{mm'}a^+_{m'},
\ee
state (\ref{eq201}) is equal to the so-called BCS state
\be\label{eq203}
|C\rangle =
\prod_{m>0}\left(1+s_m c_m \bar{a}^+_{\tilde{m}}\bar{a}^+_m\right)|0\rangle,
\ee
better known in its normalized form,
\be\label{eq204}
\frac{|C\rangle}{\langle{}C|C\rangle^{1/2}} =
\prod_{m>0}\left(u_m+s_m v_m \bar{a}^+_{\tilde{m}}\bar{a}^+_m\right)|0\rangle,
\ee
for
\be\label{eq205}
u_m = \frac{1}{\sqrt{1+c_m^2}}~~,~~v_m = \frac{c_m}{\sqrt{1+c_m^2}}~.
\ee

The Bloch-Messiah-Zumino theorem \cite{[Blo62],[Zum62]} provides the
link between the two forms of state $|C\rangle$,
Eqs.~(\ref{eq201}) and (\ref{eq203}), by stating that every complex
antisymmetric matrix can be brought by a unitary transformation into its
canonical form, i.e.,
\be\label{eq206}
(U^TCU)_{mn} = s^{\displaystyle\ast}_nc_n\delta_{m\tilde{n}},
\ee
where index $\tilde{m}$ denotes the so-called canonical partner of
the state $m$, the phase factors
$s^{\displaystyle\ast}_n$=$s^{-1}_n$=$-s^{\displaystyle\ast}_{\tilde{n}}$
have for the canonical partners opposite signs, and numbers
$c_n$=$c_{\tilde{n}}$ are real and positive.
Standard notation $m$$>$0, used in Eqs.~(\ref{eq203}) and (\ref{eq204}),
means that only one state is taken from each canonical pair. The
proof of the theorem goes through a diagonalization of the hermitian
matrix $C^+C$, that yields the unitary transformation $U$ and real,
non-negative, pairwise degenerate eigenvalues $c^2_n$.

When using states (\ref{eq201}) in applications beyond the mean-field
approximation, and in particular in the generator coordinate method (GCM)
\cite{[Hil53],[Oni66],[Won75],[RS80]}, the matrix elements and
overlaps depend on the product matrix $C^+C'$. For example, the
overlap of two states (\ref{eq201}) reads \cite{[Low55],[Oni66],[RS80]}
\be\label{eq207}
\langle{}C'|C\rangle = {\det}^{1/2}\left(1+C^+C'\right),
\ee
and the transition density matrix is given by \cite{[RS80]}
\be\label{eq208}
\rho_{mn} = \frac{\langle{}C'|a^+_n a_m|C\rangle}{\langle{}C'|C\rangle} =
\left[\left(1+C^+C'\right)^{-1}C^+C'\right]_{mn}.
\ee
It has been realized long time ago \cite{[Nee83]} that the matrix
$C^+C'$ is also pairwise degenerate, which facilitates calculation of
the phase of the overlap, otherwise ambiguous because of the square
root appearing in Eq.~(\ref{eq207}). Moreover, under certain
conditions it has been proved in Ref.~\cite{[Bur95]} that it is
enough to give up the unitarity of matrix $U$ to bring both matrices
$C$ and $C'$ {\em simultaneously} into the canonical forms analogous to
(\ref{eq206}). The same fact has later been rediscovered in
Ref.~\cite{[Don98]}, although the necessary restrictions on matrices
$C$ and $C'$ have not been recognized there.

In the present paper, I generalize results of Ref.~\cite{[Bur95]} by
deriving canonical forms of two arbitrary complex matrices $C$ and
$C'$ in a common canonical basis. These results are not restricted by
any conditions on matrices $C$ and $C'$.

Let us begin by recalling the notion of the Jordan form
(see e.g.~\cite{[Hor85]}) of an arbitrary complex matrix.
Focussing our attention on the matrix $C^+C'$, the vectors
defining its Jordan basis can be arranged in columns
of matrix $W$, and one has
   \begin{equation}\label{Ap1}
   \sum_{n}\left(C^+C'\right)_{mn}W_{ni}=\sum_j W_{mj}D_{ji},
   \end{equation}
where matrix $D$ is block diagonal (composed of the Jordan blocks).
One can attribute the number of the
block $I_i$, the length of the block $L_i$, and the number within the
block $k_i$, to every index $i$ that numbers the Jordan basis
vectors.
In this notation, $D$ has the form:
   \begin{equation}\label{eq-r2}
   D_{ji}=\delta_{I_jI_i}D^{I_i}_{k_jk_i},
   \end{equation}
where in every block matrix $D^I_{kk'}$ reads
   \begin{equation}\label{eq-r3}
    D^{I}_{kk'}=D_I\delta_{kk'}+\delta_{kk'-1},
   \end{equation}
i.e., it has a common complex number $D_I$ on the main diagonal and
the ones just above the main diagonal.

Basis vectors belonging to a given block $I$ form the so-called
Jordan series of length $L$. The series starts with the basis vector
called the series head, and ends with an eigenvector of $C^+C'$. The
whole series is uniquely determined by the series head, because the
remaining basis vectors in the series can be obtained by a repeated
action of $C^+C'$ on the series head.
The basis vectors in a given series are not unique, because a linear
combination of these vectors may give another valid series head, and
leads to the same Jordan canonical form. Explicitly, this transformation
reads
   \begin{equation}\label{eq-r5}
    W'_{mk'}=\sum_{k=1}^L W_{mk}\alpha_{kk'}=\sum_{k=1}^{k'} W_{mk}\alpha_{L-k+1},
   \end{equation}
where the transformation matrix $\alpha_{kk'}$ depends on $L$
arbitrary complex numbers $\alpha_k$ (only $\alpha_1$ must not vanish),
and has the following explicit structure:
   \be\label{eq-r6}
       \alpha_{kk'}
         = \left(\ba{lll@{~\ldots~}ll}
          \alpha_1 & \alpha_2 & \alpha_3 & \alpha_{L-1} & \alpha_L     \\
           0       & \alpha_1 & \alpha_2 & \alpha_{L-2} & \alpha_{L-1} \\
           0       &  0       & \alpha_1 & \alpha_{L-3} & \alpha_{L-2} \\
        \multicolumn{5}{c}{\dotfill}                                   \\
           0       &  0       &  0       & \alpha_1     & \alpha_2     \\
           0       &  0       &  0       &  0           & \alpha_1     \\
                 \ea\right).
   \ee
It is easy to check that matrices having this structure form a group.
It is also easy to check that vectors $W'_{mk}$ form the Jordan
series, similarly as vectors $W_{mk}$ do, and that they can replace
vectors $W_{mk}$ in the Jordan basis, giving the same matrix $D$ in
Eq.~(\ref{Ap1}).

According to the Jordan construction, the whole space in which acts
matrix $C^+C'$ splits into subspaces spanned by the Jordan series.
The number of eigenvectors of $C^+C'$ equals to the number of
different series, or to the number of Jordan blocks, and is in
general smaller than the dimension of the matrix $C^+C'$. Some
matrices (hermitian or not) can be fully diagonalized, i.e., they
have numbers of eigenvectors equal to their dimensions. This
corresponds to the case when all the Jordan series have the length
equal 1.

One calls two blocks degenerate, or two series degenerate, if they
have the same diagonal number $D_I$, {\em and} they have the same length
$L$. The latter condition is very important, because only degenerate
series defined in such a way can be mixed; this is an analogue of the
possibility to mix degenerate eigenvectors of a matrix which can be
fully diagonalized. If two series have different lengths then vectors
of a longer series cannot be admixed to those of the shorter series,
even if the series have the same diagonal number $D_I$. If the matrix
can be fully diagonalized, then all the series have length 1, the
number of eigenvectors equals to the dimension of $C^+C'$, and
transformation (\ref{eq-r5}) reduces to the possibility of
arbitrarily normalizing every eigenvector.

After these necessary preliminaries, let us proceed with presenting
the main results of the present paper.
Multiplying from the left-hand and right-hand sides the
eigen-equation (\ref{Ap1}) by $W^{-1}$, and then transposing, we
obtain that
   \begin{equation}\label{Ap2}
   C'\left(C^+{W^{-1}}^T\right)={W^{-1}}^T D^T\;,
   \end{equation}
which multiplied by $C^+$ from the left-hand side gives
   \begin{equation}\label{Ap2a}
   \left(C^+C'\right)\left(C^+{W^{-1}}^T\right)
     =\left(C^+{W^{-1}}^T\right)D^T\;.
   \end{equation}

One can see now that matrix $C^+C'$ has another equivalent set of Jordan
series, i.e.,
   \begin{equation}\label{Ap1a}
   \sum_{n}\left(C^+C'\right)_{mn}V_{ni}=\sum_j V_{mj}D_{ji},
   \end{equation}
where the new matrix of basis vectors $V$ is given by
   \begin{equation}\label{eq-r7}
    V=C^+{W^{-1}}^T J.
   \end{equation}
In every Jordan block, matrix $J$ has the ones on the skew diagonal
and zeros otherwise, i.e., $J_{kk'}$=$\delta_{k,L-k'+1}$. Hence, when
an arbitrary matrix is multiplied by $J$ from the right-hand
(left-hand) side, the order of its columns (rows) is flipped. In
particular, one obtains that $D^T$=$JDJ$.

We can now analyze cases of different degeneracies of the Jordan
blocks. The arguments given below closely follow proofs presented in
Ref.~\cite{[Bur95]}, only with the degeneracies of eigenvalues
replaced by the degeneracies of the Jordan blocks.

Let us first suppose that $C^+C'$ has a non-degenerate Jordan
block. Then, in this block the basis vectors $V$ must be connected
with the basis vectors $W$ by transformation (\ref{eq-r5}), i.e.,
   \begin{equation}\label{Ap3}
   \left(C^+{W^{-1}}^T J\right)_{mk'} = \sum_{k=1}^L W_{mk}\alpha_{kk'}.
   \end{equation}
Multiplying Eq.~(\ref{Ap3}) by
$W^{-1}$ from the left-hand side, and by $J$ from the right-hand side,
one obtains
   \begin{equation}\label{Ap4}
   \left(W^{-1}C^+{W^{-1}}^T\right)_{kk'} = \left(\alpha J\right)_{kk'}.
   \end{equation}
However, the matrix on the left-hand side of this equation is
antisymmetric, while that on the right-hand side is symmetric;
therefore, matrix $\alpha_{kk'}$ must vanish.
This
requires that, in a non-degenerate Jordan block, all vectors $V$
vanish, which contradicts Eq.~(\ref{Ap2}), unless the block
has length $L$=1 and $D_I$=0. Therefore, matrix $C^+C'$ cannot have
non-degenerate Jordan blocks apart from the subspace of $L$=1 eigenvectors
with all eigenvalues equal zero.

One can set this subspace aside, and assume from now on that $C^+C'$
is non-singular and has an even dimension. In this case, matrix
$C^+C'$ cannot have any non-degenerate Jordan block, and hence Jordan
blocks must appear in degenerate pairs. (In odd dimensions, $C^+C'$
must have at least one null eigenvalue, which can be separated, and
the remaining matrix can be treated in the even dimension).

In the present considerations, it is enough to consider only pairs of
degenerate blocks; had the higher degeneracies of the Jordan blocks
occurred, one could have considered one pair after another, and at
each step one could reduce the dimension of the problem. This is
possible here, and has not been possible when considering degenerate
eigenvalues in Ref.~\cite{[Bur95]}, because the whole space can be
separated into the subspaces corresponding to the Jordan blocks,
while it cannot be separated into subpaces corresponding to the
eigenvalues.

Let us now consider a pair of degenerate Jordan blocks, each block
having length $L$ and the common diagonal element
$D_I$=$D_{\tilde{I}}$. One can adopt here the standard notation
that originally pertains to the canonical pairs, namely, we denote
the indices of the two degenerate blocks by $I$ and $\tilde{I}$.
Similarly, indices inside these two blocks are denoted by $k$=1,
2,\ldots,$L$ and $\tilde{k}$=1, 2,\ldots,$L$, respectively. Note
that vectors in these two blocks form series, i.e., they are arranged
in a specific order; therefore a vector at a given position must be
associated with the vector at the same position in the second block.

Since for matrix $C^+C'$ two equivalent Jordan bases exist, $W$ and $V$,
vectors in series $V$ must be linear combinations of those in
series $W$.
In the pair of degenerate Jordan blocks,
this leads to the following relations between the two series:
\bnl
\left(C^+{W^{-1}}^T J\right)_{mk'}                           \label{eq-r9a}
             &=&\!\sum_{k        =1}^L W_{mk        }\alpha_{kk'}
                + \sum_{\tilde{k}=1}^L W_{m\tilde{k}} \beta_{kk'},\\
\left(C^+{W^{-1}}^T J\right)_{m\tilde{k}'}                   \label{eq-r9b}
             &=&\!\sum_{k        =1}^L W_{mk        }\gamma_{kk'}
                + \sum_{\tilde{k}=1}^L W_{m\tilde{k}}\epsilon_{kk'},
\enl
All the four matrices $\alpha$, $\beta$, $\gamma$, and $\epsilon$
have the same structure (\ref{eq-r6}). One may now proceed with
multiplying Eqs.~(\ref{eq-r9a}) and (\ref{eq-r9b}) from the left-hand
side either by $W^{-1}_{km}$ or by  $W^{-1}_{\tilde{k}m}$, and
from the right-hand side by $J$. Since all matrices $\alpha{J}$,
$\beta{J}$, $\gamma{J}$, and $\epsilon{J}$ are symmetric, one then
obtains that $\alpha$=$\epsilon$=0 and $\gamma$=$-\beta$.

Therefore, the canonical form of the $C^+$ matrix reads
   \begin{equation}\label{Ap10a}
   \left({W^{-1}} C^+ {W^{-1}}^T\right)_{ji}
     =s_{I_j}C^{I_j\,+}_{k_jk_i}
                        \delta_{I_j\tilde{I}_i},
   \end{equation}
where the symmetric matrix $C^{I}$=$C^{\tilde{I}}$ occupies the
off-diagonal part in every pair of the degenerate Jordan blocks, and
   \begin{equation}\label{eq-r10}
   C^{I} =   s_I\beta^{\displaystyle\ast} J ,
   \end{equation}
for $\beta$ having form (\ref{eq-r6}). Following the standard
notation, we have defined the phase factors $s_I$=$-s_{\tilde{I}}$
in such a way that the skew-diagonal matrix elements of $C^I$ (that
are all equal one to another) are real and positive, i.e.,
$C^I_{k,L-k+1}$$>$0.

Since the canonical basis of $C^+$ is the same as the Jordan
basis of $C^+C'$, matrix $C'$ must in the very same basis assume
an analogous canonical form:
   \begin{equation}\label{Ap11a}
   \left(W^T C' W\right)_{ji}
     = s_{I_i}^{\displaystyle\ast} C'^{I_i}_{k_jk_i}
                        \delta_{I_j\tilde{I}_i},
   \end{equation}
where the symmetric matrix $C'^{I}$=$C'^{\tilde{I}}$ reads
   \begin{equation}\label{eq-r11}
    C'^{I} =  s_I^{\displaystyle\ast}J \beta'^{\displaystyle\ast}
   \end{equation}
and $\beta'$ has also form (\ref{eq-r6}). Finally, in order to
satisfy Eq.~(\ref{Ap1}) matrices $\beta$ and $\beta'$ must obey the
following condition:
   \begin{equation}\label{eq-r12}
   \beta  \beta'^{\displaystyle\ast}  = D .
   \end{equation}
This leaves us still some freedom in the choice of the canonical
basis, because any solution of Eq.~(\ref{eq-r12}) gives one valid
canonical form. Two obvious choices are, for example,
$\beta$=$D$ and $\beta'^{\displaystyle\ast}$=$I$
or
$\beta$=$\sqrt{D}$ and $\beta'^{\displaystyle\ast}$=$\sqrt{D}$,
where any one of the possible branches of the matrix square root
can be taken.

Equations (\ref{Ap10a}) and  (\ref{Ap11a}) complete the proof of the
canonical forms of two arbitrary complex antisymmetric matrices
$C$ and $C'$. Both these matrices can be {\em simultaneously}
transformed by matrix $W$ (in general non-unitary) into the
block-diagonal forms with non-zero elements only between pairs of
degenerate Jordan blocks.

Needless to say, whenever matrix $C^+C'$ can be fully diagonalized,
which was the case in Ref.~\cite{[Bur95]}, both matrices $C$ and $C'$
acquire in the canonical basis the standard canonical forms
analogous to Eq.~(\ref{eq206}), i.e.,
   \begin{equation}\label{Ap10}
   \left({W^{-1}} C^+ {W^{-1}}^T\right)_{ji}
     =s_j c_{j}^{\displaystyle\ast}\delta_{j\tilde\imath}\;,
   \end{equation}
and
   \begin{equation}\label{Ap11}
   \left(W^T C' W\right)_{ji}
     =s_i^{\displaystyle\ast} c'_{i}\delta_{j\tilde\imath}\;,
   \end{equation}
where $c_{i}^{\displaystyle\ast}c'_{i}$=$D_i$.

In Ref.~\cite{[Rob99]} it was noticed that an incorrect conjecture
was formulated in Ref.~\cite{[Bur95]}, namely, the conjecture that
the simple forms of Eqs.~(\ref{Ap10}) and  (\ref{Ap11}) can always be
achieved. In the present study we have seen that these simple forms
occur only when matrix $C^+C'$ can be fully diagonalized. In
fact, this is the case which occurs most often in applications.
Therefore, let us now discuss conditions for the full diagonalization
of $C^+C'$.

In the applications given in Ref.~\cite{[Bur95]}, the full
diagonalization of matrix $C^+C'$ was secured by using a model in
which matrices $C$ were time-even,
\be\label{eq209}
C^+=U_TC^TU_T^T,
\ee
and the Hermitian and time-even matrices $\tilde{C}$ defined by
\be\label{eq210}
\tilde{C}= -U_TC
\ee
were positive definite. In these equations, $U_T$ is a unitary and
antisymmetric matrix, $U_T^+$=$U_T^{-1}$=$-U_T^{\displaystyle\ast}$.
The positive definiteness of matrices $\tilde{C}$ was in
\cite{[Bur95]} guaranteed by a special form of matrices $C$. In that
study, the GCM states were constructed within the SCEM model
\cite{[Dob90],[Dob91]}, and therefore matrices $C$ had the form shown
in Eq.~(2.13) of \cite{[Bur95]}. Therefore, the corresponding
$\tilde{C}$ matrices were all equal to exponents of hermitian
matrices, and hence trivially positive definite.

In the general presentation of the present paper, conditions
(\ref{eq209}) and (\ref{eq210}) can be formulated as follows: If
there exist a unitary antisymmetric matrix $U_T$ such that
Eq.~(\ref{eq209}) holds for $C$ and $C'$, and at the same time at
least $\tilde{C}$ or $\tilde{C'}$ is positive-definite, then matrix
$C^+C'$ can be fully diagonalized, and the simple canonical forms
(\ref{Ap10}) and  (\ref{Ap11}) exist. The proof of this statement has
been given in Ref.~\cite{[Bur95]} (Appendix C), and will not be
repeated here.

The positive definiteness of $\tilde{C}$ is a required condition,
because the hermitian square-root of $\tilde{C}$ must exist.
Unfortunately, this condition cannot be released, i.e., if both
$\tilde{C}$ and $\tilde{C'}$ are not positive definite, it may happen
that matrix $C^+C'$ cannot be fully diagonalized.
An example of such a situation is provided by the following
two 4$\times$4 matrices:
   \begin{equation}\label{eq-r19}
   C = \left(\ba{cc} 0      &  A \\
                   -A^T     &  0         \ea\right)  \mbox{~~~~and~~~~}
   C'= \left(\ba{cc} 0      &  A'\\
                   -A'^T    &  0         \ea\right) ,
   \end{equation}
where the two-dimensional matrices $A$ and $A'$ read
   \begin{equation}\label{eq-r20}
   A = \left(\ba{cc} 1      &  a \\
                     a^{\displaystyle\ast}    &  0         \ea\right)  \mbox{~~~~and~~~~}
   A'= \left(\ba{cc} 0      &  1 \\
                     1      &  0         \ea\right) .
   \end{equation}
For the standard time-reversal matrix $U_T$ given by
   \begin{equation}\label{eq-r21}
   U_T = \left(\ba{cc} 0      &  I \\
                      -I      &  0         \ea\right) \mbox{~~~~one has~~~~}
   \tilde{C} = \left(\ba{cc} A^T    &  0 \\
                             0      &  A         \ea\right) ,
   \end{equation}
and $\tilde{C}'$ has the same form. Neither $\tilde{C}$
(for $a$$\neq$0)
nor $\tilde{C}'$ is positive definite,
and the $C^+C'$=$\tilde{C}\tilde{C}'$ matrix,
   \begin{equation}\label{eq-r23}
    C^+C' =
       \left(\ba{cc} (AA')^{\displaystyle\ast}    &  0 \\
                      0     &  AA'       \ea\right)  \mbox{~~for~~}
          AA'= \left(\ba{cc} a      &  1 \\
                             0      &  a^{\displaystyle\ast}       \ea\right) ,
   \end{equation}
cannot be fully diagonalized, unless $a$$\neq$$a^{\displaystyle\ast}$.

However, for any small but non-zero imaginary part of $a$,
matrix $C^+C'$ {\em can} be fully diagonalized. Therefore, this
example also shows that the positive definiteness of (time-even)
matrices $\tilde{C}$ or $\tilde{C}'$ is only a sufficient condition
for the full diagonalization of
$C^+C'$=$\tilde{C}\tilde{C}'$, but it is not necessary.
Moreover, it is clear that matrix $C^+C'$ cannot be diagonalized for
$a$=$a^{\displaystyle\ast}$, because in the limit of
$\Im{a}$$\longrightarrow$0 two eigenvectors of $C^+C'$ become
parallel. This illustrates the difficulty of diagonalizing $C^+C'$
numerically for small values of $\Im{a}$; the task is then bound to
become ill-conditioned.

In the GCM, matrices $C$ are most often obtained from
solutions of the Hartree-Fock-Bogoliubov (HFB) or Hartree-Fock+BCS
\cite{[RS80]} equations for time-even states. In these cases,
matrices $\tilde{C}$ are diagonal in the HFB or BCS canonical bases
\cite{[RS80]} (composed of pairs of time-reversed states), and their
eigenvalues are equal to $v_m/u_m$=$c_m$, where $v_m$ and $u_m$ are
the standard quasiparticle amplitudes of Eq.~(\ref{eq205}). Here, the
canonical pairs are defined by the time reversal, and therefore the
eigenvalues $c_m$, can have, in principle, arbitrary signs.

However, in the BCS method (with a constant gap parameter $\Delta$)
all these quasiparticle amplitudes are positive, and hence all the
resulting $\tilde{C}$ matrices are positive definite, thus
fulfilling the sufficient condition for the full diagonalization of
$C^+C'$=$\tilde{C}\tilde{C}'$. In fact, quasiparticle
amplitudes of different signs rarely occur in nuclear physics
applications, cf.~Ref.~\cite{[Hah91]}. This is so, because typical
pairing forces couple the time-reversed states, and, in general, are
always attractive. This shows that the Jordan structures discussed
here cannot be expected to be frequently encountered, and most often
one will deal with the standard canonical forms of Eqs.~(\ref{Ap10})
and  (\ref{Ap11}), in which the only non-zero matrix elements are
adjacent to the main diagonal.

In summary, I have shown how to extend the results of
Ref.~\cite{[Bur95]} in order to construct canonical basis in which
two arbitrary complex antisymmetric matrices $C$ and $C'$ acquire
simultaneously canonical forms. This construction completes the
generalization of the classic Bloch-Messiah-Zumino theorem to the
case of non-diagonal matrix elements calculated between fermion-pair
condensates.

The critical reading of the manuscript by S.G.~Rohozi\'nski
is gratefully appreciated.
This research was supported by the Polish Committee for
Scientific Research, Contract No.~2~P03B~040~14.


\begin{thebibliography}{10}

\bibitem{[Bar57]}
{J. Bardeen, L.N. Cooper, and J.R. Schrieffer, Phys. Rev. {\bf 108}, 1175
  (1957)}.

\bibitem{[Bog]}
{N.N. Bogoliubov, Sov. Phys. JETP {\bf 7}, 41 (1958)};
%
{Sov. Phys. Usp. {\bf 2}, 236 (1959)};
%
{Usp. Fiz. Nauk. {\bf 67}, 549 (1959)}.

\bibitem{[Blo62]}
{C. Bloch and A. Messiah, Nucl. Phys. {\bf 39}, 95 (1962)}.

\bibitem{[Zum62]}
{B. Zumino, J. Math. Phys. {\bf 3}, 1055 (1962)}.

\bibitem{[Hil53]}
{D.L. Hill and J.A. Wheeler, Phys. Rev. {\bf 89}, 1102 (1953)}.

\bibitem{[Oni66]}
{N. Onishi and S. Yoshida, Nucl. Phys. {\bf 80}, 367 (1966)}.

\bibitem{[Won75]}
{C. W. Wong, Phys. Rep. {\bf 15C}, 283 (1975)}.

\bibitem{[RS80]}
{P. Ring and P. Schuck, {\sl The Nuclear Many-Body Problem} (Springer-Verlag,
  Berlin, 1980)}.

\bibitem{[Low55]}
{P.O. L\"owdin, Phys. Rev. {\bf 97}, 1474 (1955)}.

\bibitem{[Nee83]}
{K. Neerg{\aa}rd and E. W\"ust, Nucl. Phys. {\bf A402}, 311 (1983)}.

\bibitem{[Bur95]}
{K. Burzy\'nski and J. Dobaczewski, Phys. Rev. {\bf C51}, 1825 (1995)}.

\bibitem{[Don98]}
{F. D\"onau, Phys. Rev. {\bf C58}, 872 (1998)}.

\bibitem{[Hor85]}
{R.A. Horn and C.R. Johnson, {\sl Matrix Analysis} (Cambridge University Press,
  Cambridge, 1985)}.

\bibitem{[Rob99]}
{L.M Robledo and J.L. Egido, unpublished}.

\bibitem{[Dob90]}
{J. Dobaczewski, Phys. Lett. {\bf 241B}, 289 (1990)}.

\bibitem{[Dob91]}
{J. Dobaczewski and S.G.~Rohozi\'nski, {\it Proceedings of the 3rd
  International Spring Seminar on Nuclear Physics}, ed. A.~Covello (World
  Scientific, Singapore, 1991) p. 351}.

\bibitem{[Hah91]}
{F.J.W. Hahne and P. Ring, Phys. Lett. B {\bf 259}, 7 (1991)}.

\end{thebibliography}

\end{document}